O.M. Oshakuade* and O.E. Awe

# Determination of bulk and surface properties of liquid Bi-Sn alloys using an improved quasi-lattice theory

**Abstract:** The thermodynamic properties of Bi-Sn were studied at 600 and 900K using a quasi-lattice theory. After successful fitting of Gibbs free energies of mixing and thermodynamic activities, the fitting parameters were used to investigate the enthalpy of mixing, the entropy of mixing, concentration fluctuations, Warren-Cowley short range order parameter, surface concentrations and surface tensions of the binary systems. Positive and symmetrically shaped enthalpies of mixing were observed in all composition range, while negative excess entropies of mixing were observed. Bi-Sn showed a slight preference for like-atoms as nearest neighbours in all composition range. The nature of atomic order in Bi-Sn at 600 and 900K appeared similar. The highest tendency for homocoordination exists at composition where mole fraction of Bi is about 40%. It was also observed that Bi (whose surface tension is lower than that of Sn) has the highest surface enrichment in the Bi-Sn systems. Unlike many previous applications of the quasi-lattice theory where constant values were used to approximate coordination numbers, temperature and composition-dependent coordination numbers were applied in this work.

## 1 Introduction

Before the last two decades, the eutectic Pb-Sn alloy has been the major soldering material in the electronics industry. However, the toxic nature of Pb has led to the restriction in the use of Pb-based materials, hence, the need for substitutes [1]. Bi-Sn and Bi-Sn-based ternary systems have been identified as Pb-free solder candidates, due to the low liquidus temperature of Bi-Sn [2]. Moreover, the determination of a good solder candidate requires several other assessments, such as wettability, solidification behaviour, solder joint strength, safety to the environment, electrical conductivity and chemical stability [3,4]. Some researchers expect ternary and multicomponent alloys to be better solder candidates because they possess lower liquidus temperatures than their binary counterparts [5]. Notwithstanding, the understanding of the thermodynamics of binary alloy systems is important in the study of their ternary and multicomponent systems [5–7].

The thermodynamic properties of the Bi-Sn binary system have been studied both experimentally and theoretically [2,8–10]. Some scientists have also worked on Bi-Sn-based ternary and multicomponent systems [4,8,9,11,12]. Experimentally-obtained thermodynamic data (Gibbs free energies of mixing, entropies and enthalpies of mixing, thermodynamic activities) of Bi-Sn for 600K is available in binary alloy repository [10]. Also, Katayama et al. [8] have studied the Bi-Sn and Bi-Sb-Sn systems at 900K via EMF method. However, a literature search on the nature of atomic order and surface properties of Bi-Sn shows that research work is deficient in that area.

Amongst several existing thermodynamic models, the Quasi-lattice theory (QLT) [13] is preferred for its ability to describe alloy systems in terms of the relative bonding energies between atom pairs and success in the study of concentration-concentration fluctuations in the long-wavelength limit. Unlike the widely used approximations of the Complex formation model by Bhatia and Hargrove [14] (where coordination number is either two or infinity), QLT [13] depends on more physically reasonable coordination numbers. In many previous applications of QLT [13], constants have been used to approximate coordination number.

In this work, the bulk and surface properties of Bi-Sn is studied using an improved quasi-lattice theory, introduced in our previous work [15], where more physically realistic coordination numbers have been incorporated into the popular quasi-lattice theory [13]. A model for estimating temperature-dependent coordination numbers of pure liquid metals, developed by Tao [16], is adopted for the estimation of nature and temperature-dependent coordination numbers in this work. Bi-Sn is studied at 600 and 900K to provide information on the temperature dependence of the bulk and surface properties, and also help in Bi-Sb-Sn studies via QLT, in future. This work will contribute to the knowledge on the

Department of Physics, University of Ibadan, Ibadan, Nigeria.
Corresponding author. Tel.: +2348038557420; E-mail: om.oshakuade@ui.edu.ng, gbengaoshaks@yahoo.com

thermodynamic properties of the Bi-Sn system which is important in understanding its process metallurgy and further materials development.

The next section gives details on QLT, its improvement, and application to Gibbs free energy of mixing, thermodynamic activity, enthalpy of formation, entropy of formation, concentration fluctuations, Warren-Cowley chemical short range order parameter and surface properties calculation. Section 3 contains results and discussions on the computed thermodynamic properties while Section 4 contains summary and conclusions.

## 2 Theoretical framework

### 2.1 Average coordination number $(\hat{Z})$

The average coordination number of liquid alloys, $\hat{Z}$, is depends on the nature, composition and temperature of the system [16–23]. Equation (1) is introduced to account for the composition dependent nature of Coordination number ($Z$) in liquid alloys [24,25]. Where $Z_i$ is the coordination number of pure element $i$ and $c_i$ is the molar concentration of component $i$ in the system.

$$\hat{Z} = \sum_{i=1}^{2} Z_i c_i \quad (1)$$

The model of Tao [16], which estimates temperature-dependent coordination numbers for pure liquid metals ($Z_i$) from their physical properties, defined in Eq. (2), is used in this work.

$$Z_i = \frac{4\sqrt{2\pi}}{3} \left( \frac{r_{mi}^3 - r_{0i}^3}{r_{mi} - r_{0i}} \right) \frac{0.6022 \, r_{mi}}{V_{mi}} \exp\left( \frac{\Delta H_{mi}(T_{mi} - T)}{Z_c R T T_{mi}} \right) \quad (2)$$

Where $T$ is the temperature of the system under study, $\Delta H_{mi}$ and $T_{mi}$ are the melting enthalpy and melting temperature of component $i$, respectively, $Z_c$ is the close-packed coordination number of magnitude 12, $V_{mi}$ is the molar volume of component $i$ at $T$, $r_{0i}$ and $r_{mi}$ are the beginning and first peak values of radial distance distribution function near the melting point of liquid metal $i$, respectively. The relevant parameters used in this work to estimate $Z_i$ are presented in **Tab. 1** [26].

Expanding Eqs. (1) and (2) for Bi-Sn binary system gives Eq. (3), an expression for average coordination number that is dependent on nature, composition and temperature.

$$\hat{Z}(c_{Bi}, T) = \frac{4 \times 0.6022 \sqrt{2\pi}}{3} \left[ \begin{array}{l} \left( \dfrac{r_{mBi}^3 - r_{0Bi}^3}{r_{mBi} - r_{0Bi}} \right) \dfrac{r_{mBi}}{V_{mBi}} \exp\left( \dfrac{\Delta H_{mBi}(T_{mBi} - T)}{Z_c R T T_{mBi}} \right) c_{Bi} \\ + \left( \dfrac{r_{mSn}^3 - r_{0Sn}^3}{r_{mSn} - r_{0Sn}} \right) \dfrac{r_{mSn}}{V_{mSn}} \exp\left( \dfrac{\Delta H_{mSn}(T_{mSn} - T)}{Z_c R T T_{mSn}} \right) (1 - c_{Bi}) \end{array} \right] \quad (3)$$

Subscripts Bi & Sn in Eq. (3) corresponds to the two components (Bi and Sn) in the Bi-Sn binary system.

The Average coordination number $(\hat{Z})$, defined in Eq. (3), is adopted in this work to apply quasi-lattice theory to the study of Bi-Sn systems.

**Tab. 1:** Calculated $Z_i$ of each component and essential parameters related to its calculation.

| Metal $i$ | $V_{mi}$ [a] (cm³/mol) | $T_{mi}$ [a] (K) | $\Delta H_{mi}$ [a] (kJ/mol) | $r_{mi}$ [a,b] | $r_{0i}$ [a,b] | $Z_i (T)$ [c] 600K | 900K |
|---|---|---|---|---|---|---|---|
| Bi | 20.80 [1 + 0.000117 ($T$ - 544)] | 544 | 10.88 | 3.34 | 2.78 | 8.8770 | 8.0736 |
| Sn | 17.00 [1 + 0.000087 ($T$ - 505)] | 505 | 7.07 | 3.14 | 2.68 | 9.1799 | 8.6028 |

[a] These parameters were obtained from [26]
[b] Unit is $10^{-8}$ cm
[c] The various values for different temperatures were obtained using Eq. (2)

## 2.2 Quasi-lattice theory for compound forming liquid alloys (QLT)

Following the assumptions by Bhatia and Singh [13] on QLT, it can be deduced that $N_{Bi}$ moles of Bi and $N_{Sn}$ moles of Sn will mix to form a pseudo-ternary mixture comprising of $n_1$ moles of unassociated Bi atoms (specie 1), $n_2$ moles of unassociated Sn atoms (specie 2) and $n_3$ moles of $Bi_\mu Sn_\nu$ complex (specie 3). Where $\mu$ and $\nu$ are small integers obtained from stoichiometric information while $c_{Bi}$ and $c_{Sn}$ are the molar concentrations of Bi & Sn, respectively ($c_{Bi} + c_{Sn} = 1$). If the total number of moles in the mixture is $N$, then Eqs. (4) - (8) can be equivalent to the laws of conservation of matter. The average coordination number $(\hat{Z})$ described in Section 2.1 is used in this section to replace the nominal $Z$, which is common in previous works on QLT.

$$N_{Bi} + N_{Sn} = N \quad (4)$$

$$N_{Bi} = N c_{Bi} \quad (5)$$

$$N_{Sn} = N - N_{Bi} = N - N c_{Bi} = N(1 - c_{Bi}) \quad (6)$$

$$n_1 = N c_{Bi} - \mu n_3 \quad (7)$$

$$n_2 = N(1 - c_{Bi}) - \nu n_3 \quad (8)$$

The expressions for $G_M$ given as Eq. (9) can be written as Eq. (10) [27,14,28].

$$G_M = G - N c_{Bi} G_{Bi}^{(0)} - N c_{Sn} G_{Sn}^{(0)} \quad (9)$$

$$G_M = -n_3 g + G' \quad (10)$$

Where $G_i^{(0)}$ is the chemical potential for specie $i$ in the mixture, $G$ is the free energy of the mixture, $g$ is the free energy of formation of chemical complex, the term "$-n_3 g$" denotes the lowering of free energy as a result of chemical complex formation, $G'$ is the free energy of mixing of the pseudo-ternary mixture assumed to interact weakly, and $G'$ depends on the underlying relevant theory being applied to study weakly interacting mixtures. Generally, $g$ and $G'$ is defined as Eqs. (11) and (12).

$$g = \mu G_{Bi}^{(0)} + \nu G_{Sn}^{(0)} - G_{BiSn}^{(0)} \quad (11)$$

$$G' = G - \left( n_1 G_{Bi}^{(0)} + n_2 G_{Sn}^{(0)} + n_3 G_{BiSn}^{(0)} \right) \quad (12)$$

For an ideal mixture, $G'$ can be defined as Eq. (13), where $R$ and $T$ are the molar gas constant and temperature, respectively.

$$G' = RT\left(n_1 \ln \frac{n_1}{n_1+n_2+n_3} + n_2 \ln \frac{n_2}{n_1+n_2+n_3} + n_3 \ln \frac{n_3}{n_1+n_2+n_3}\right) \quad (13)$$

Assuming only one type of complex is formed, the QLT by Bhatia and Singh [13] gives the expression for $G_M$ as Eq. (14), where $\zeta$ & $\aleph$ are defined in Eqs. (15) & (16), respectively.

$$G_M = -n_3 g + RT\left[\begin{array}{c} n_1 \ln \frac{n_1}{N} + n_2 \ln \frac{n_2}{N} + n_3 \ln \frac{(\mu+\nu)n_3}{N} \\ -\frac{1}{2}\hat{Z}n_3(\mu+\nu-\zeta)\ln \frac{\mu+\nu}{\mu+\nu-\zeta} - \frac{1}{2}\hat{Z}\aleph \ln \frac{\aleph}{N} \end{array}\right] + \frac{1}{\aleph}\left(\begin{array}{c} \frac{n_1 n_2}{N} v_{12} + \frac{n_1 n_3}{N} v_{13} \\ + \frac{n_2 n_3}{N} v_{23} \end{array}\right) \quad (14)$$

$$\zeta = \frac{2(\mu+\nu-1)}{\hat{Z}} \quad (15)$$

$$\aleph = n_1 + n_2 + \left(\mu+\nu - \frac{2(\mu+\nu-1)}{\hat{Z}}\right)n_3 = N - \frac{2(\mu+\nu-1)}{\hat{Z}}n_3 = N - \zeta n_3 \quad (16)$$

The $v_{ij}$s give information on the inter-specie energetics of species $i$ and $j$ in the pseudo ternary mixture. Interaction parameters $g$, $v_{12}$, $v_{13}$ and $v_{23}$ are determined by fitting, until Eq. (14) conforms to experimental thermodynamic data. The equilibrium condition of $n_3$ at a specified temperature is given in Eq. (17) [13].

$$\left(\frac{\partial G_M}{\partial n_3}\right)_{T,P,N,c_1'} = 0 \quad (17)$$

When Eq. (14) is substituted for $G_M$ in Eq. (17), it gives Eq. (18) (where Q is defined in Eq. (19)).

$$n_1^\mu n_2^\nu = (\mu+\nu)n_3 e^{(Q-g/RT)} \aleph^{\mu+\nu-1} \quad (18)$$

$$Q = \frac{1}{\aleph RT}\left(\begin{array}{c} -(\nu n_1 + \mu n_2)v_{12} + (n_1 - \mu n_3)v_{13} \\ +(n_2 - \nu n_3)v_{23} + \frac{\zeta}{\aleph}(n_1 n_2 v_{12} + n_1 n_3 v_{13} + n_2 n_3 v_{23}) \end{array}\right) - \frac{1}{2}\hat{Z}(\mu+\nu-\zeta)\ln \frac{\mu+\nu}{\mu+\nu-\zeta} \quad (19)$$

The directly observed thermodynamic activity can be derived from thermodynamic function in Eq. (20), where $i$ refers to Bi or Sn [13,29]. Simplification of Eq. (20) for components Bi and Sn while considering the composition-dependent nature of $\hat{Z}$ gives Eqs. (21) and (22). Where $Z_{Bi}$ and $Z_{Sn}$ are the coordination numbers of pure liquid Bi and Sn at the specified $T$, respectively.

$$\ln a_i = \frac{1}{RT}\left(\frac{\partial G_M}{\partial N_i}\right)_{T,P,N} \quad (20)$$

$$\ln a_{Bi} = \ln\left(\frac{n_1}{N}\right) - \frac{1}{2}\hat{Z}\ln\left(\frac{\aleph}{N}\right) + \frac{n_2 v_{12} + n_3 v_{13}}{\aleph RT} - \frac{n_1 n_2 v_{12} + n_1 n_3 v_{13} + n_2 n_3 v_{23}}{\aleph^2 RT}$$
$$- \frac{Z_{Bi}-\hat{Z}}{2N}\left[N\ln\left(\frac{\aleph}{N}\right) + n_3(\mu+\nu)\ln\left(\frac{\mu+\nu}{\mu+\nu-\zeta}\right)\right] \quad (21)$$
$$- \frac{n_3 \zeta}{\aleph^2 RT}\frac{Z_{Bi}-\hat{Z}}{\hat{Z} N}(n_1 n_2 v_{12} + n_1 n_3 v_{13} + n_2 n_3 v_{23})$$

$$\ln a_{Sn} = \ln\left(\frac{n_2}{N}\right) - \frac{1}{2}\hat{Z}\ln\left(\frac{\aleph}{N}\right) + \frac{n_1 V_{12} + n_3 V_{23}}{\aleph RT} - \frac{n_1 n_2 V_{12} + n_1 n_3 V_{13} + n_2 n_3 V_{23}}{\aleph^2 RT}$$
$$- \frac{Z_{Sn} - \hat{Z}}{2N}\left[N\ln\left(\frac{\aleph}{N}\right) + n_3(\mu+\nu)\ln\left(\frac{\mu+\nu}{\mu+\nu-\zeta}\right)\right] \quad (22)$$
$$- \frac{n_3 \zeta}{\aleph^2 RT} \frac{Z_{Sn} - \hat{Z}}{\hat{Z} N}(n_1 n_2 V_{12} + n_1 n_3 V_{13} + n_2 n_3 V_{23})$$

Equation (18) is solved numerically to obtain $n_3$ at each concentration of study, having set the interaction parameters to convenient start values. The corresponding $n_1$ and $n_2$ for each $n_3$ are obtained from Eqs. (7) and (8). Thereafter, $n_1$, $n_2$ and $n_3$ are used in Eqs. (14), (21) & (22) to compute $G_M$ and $a_i$s. The computed $G_M$ and $a_i$s are then compared with experimental $G_M$ and $a_i$s, and adjustment of interaction parameters ($g$, $v_{12}$, $v_{13}$, $v_{23}$) is made in order to obtain a good fit. The final set of interaction parameters that give satisfactory agreement between experimental and computed data are then used to compute other thermodynamic quantities.

Mean absolute percentage error ($Er_i$) estimation, as defined in Eq. (23), was used to measure the agreements between two sets of data [30].

$$Er_i = \pm \frac{100}{t} \sum_{i=1}^{t}\left[\frac{d_{i,ex} - d_{i,pr}}{d_{i,ex}}\right] \quad (23)$$

Where, $d_{i,ex}$ and $d_{i,pr}$ are the existing and predicted values, respectively, while $t$ is the size of data set.

The assumed complex obtained from stoichiometric information [8,10] is $Bi_1Sn_1$. **Table 2** shows the interaction parameters obtained for Bi-Sn at 600 and 900K and the corresponding mean absolute percentage error in QLT computation. Bi-Sn can be described as weakly interacting because their values of $g/RT$ is small compared to strongly interacting systems like Mg-Bi, Tl-Te, K-Te and Na-Sn with interaction parameters ranging from 16.7 to 47.8 [14,31,32]. The negative $v_{ij}$s imply attractive interactions between the species. The theoretically fitted $G_M$ and $a_i$s, with corresponding experimentally observed values [8,10], for Bi-Sn at both temperatures are plotted in **Fig. 1** and **Fig. 2**. The small $Er_i$ values in **Tab. 2** and the plots presented in **Fig. 1** and **Fig. 2** show that the interaction parameters for both temperatures give good fits. Hence, the interaction parameters, listed in **Tab. 2**, will be used in further thermodynamic studies in this work.

**Tab. 2:** Interaction parameters for each binary system and their corresponding mean absolute percentage error.

| System | $\mu$ | $\nu$ | $\frac{g}{RT}$ | $\frac{v_{12}}{RT}$ | $\frac{v_{13}}{RT}$ | $\frac{v_{23}}{RT}$ | $Er_i$ (%) $G_M$ | $a_i$s |
|---|---|---|---|---|---|---|---|---|
| Bi-Sn at 600K | 1 | 1 | -1.8 | 0.5 | -0.2 | 0.3 | 0.19 | 0.25 |
| Bi-Sn at 900K | 1 | 1 | -1.8 | 0.5 | -0.2 | 0.2 | 0.83 | 1.23 |

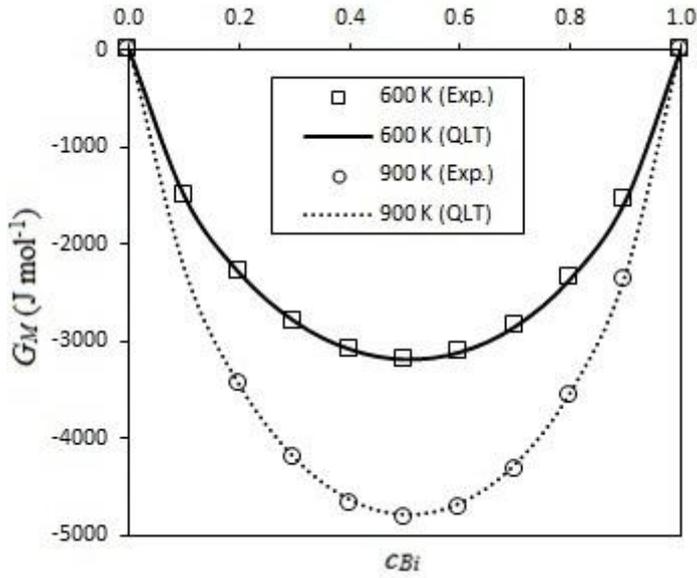

**Fig. 1:** GM in Bi-Sn at 600 & 900K.

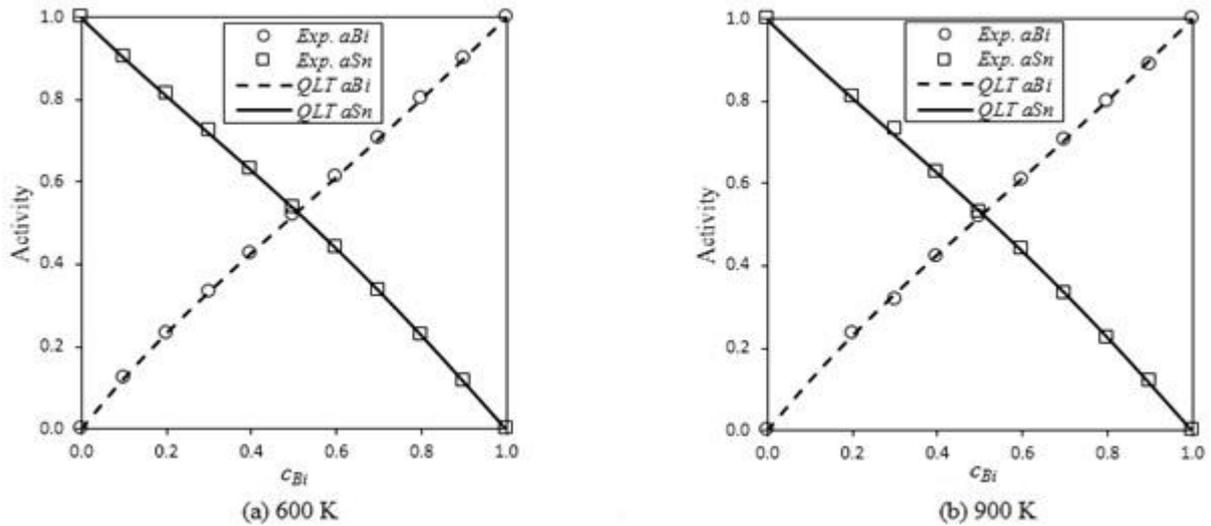

**Fig. 2:** Activities of all components in Bi-Sn at 600 & 900K.

### 2.2.1 Enthalpy and entropy of mixing

To study the enthalpy and entropy of mixing within the quasi-lattice theory, the knowledge of the partial differentials of the interaction parameters with respect to temperature is required. The enthalpy of mixing, $H_M$, can be defined from thermodynamic relations [28,33] as given in Eq. (24). Differentiating $G_M$, Eq. (14), which contains the temperature-dependent $g$, $v_{ij}s$ and $\hat{Z}$, and substituting in Eq. (24) gives Eq. (25).

$$H_M = G_M - T\left(\frac{\partial G_M}{\partial T}\right)_P \quad (24)$$

$$H_M = -n_3\left(g - T\frac{\partial g}{\partial T}\right) + \frac{1}{2}\frac{\partial \hat{Z}}{\partial T}RT^2\left(n_3(\mu+\nu)\ln\frac{\mu+\nu}{\mu+\nu-\zeta} + N\ln\frac{\aleph}{N}\right)$$
$$+ \frac{1}{\aleph}\sum\sum_{i<j}^{3}n_i n_j\left(v_{ij} - T\frac{\partial v_{ij}}{\partial T}\right) + T\frac{\partial \hat{Z}}{\partial T}\frac{n_3\zeta}{\aleph^2\hat{Z}}\sum\sum_{i<j}^{3}n_i n_j v_{ij} \quad (25)$$

$$S_M = \frac{H_M - G_M}{T} \quad (26)$$

$$S_M = n_3\frac{\partial g}{\partial T} - R\left(\begin{array}{c} n_1\ln\frac{n_1}{N} + n_2\ln\frac{n_2}{N} + n_3\ln\frac{(\mu+\nu)n_3}{N} \\ -\frac{1}{2}\hat{Z}n_3(\mu+\nu-\zeta)\ln\frac{\mu+\nu}{\mu+\nu-\zeta} - \frac{1}{2}\hat{Z}\aleph\ln\frac{\aleph}{N} \end{array}\right)$$
$$+ \frac{1}{2}\frac{\partial \hat{Z}}{\partial T}RT\left(n_3(\mu+\nu)\ln\frac{\mu+\nu}{\mu+\nu-\zeta} + N\ln\frac{\aleph}{N}\right) \quad (27)$$
$$- \frac{1}{\aleph}\sum\sum_{i<j}^{3}n_i n_j\frac{\partial v_{ij}}{\partial T} + \frac{\partial \hat{Z}}{\partial T}\frac{n_3\zeta}{\aleph^2\hat{Z}}\sum\sum_{i<j}^{3}n_i n_j v_{ij}$$

$$\frac{\partial \hat{Z}}{\partial T} = -Z_{Bi}c_{Bi}\left(\frac{\Delta H_{mBi}}{Z_c RT^2} + \frac{1}{T - T_{mBi} + 1}\right) - Z_{Sn}c_{Sn}\left(\frac{\Delta H_{mSn}}{Z_c RT^2} + \frac{1}{T - T_{mSn} + 1}\right) \quad (28)$$

The entropy of mixing, $S_M$, is defined in Eq. (26) [28,33]. The expression for $S_M$ is simplified and given in Eq. (27). Equation (27) is obtained by substituting Eqs. (14) and (25) for $G_M$ and $H_M$, respectively, in Eq. (26). The partial differential of the average coordination number with respect to system temperature $(\partial \hat{Z}/\partial T)$ in Eqs. (24) and (27) is defined in (28). The $\partial \hat{Z}/\partial T$ in Eq. (28) is based on the model of Tao [16], which is used in this work to compute coordination numbers of pure liquid metals. Since the interaction parameters are temperature dependent, the partial differentials: $\partial g/\partial T$, $\partial v_{12}/\partial T$, $\partial v_{13}/\partial T$ and $\partial v_{23}/\partial T$, were fitted to conform with experimental $H_M$ and $S_M$, for Bi-Sn at 600K. Excess entropy of mixing ($S_M^{xs}$) was estimated from real and ideal entropies of mixing, using thermodynamic relations given in Eq. (29) [28,33].

$$S_M^{xs} = S_M + NR\sum c_i \ln c_i \quad (29)$$

### 2.2.2 Concentration-concentration fluctuations in the long-wavelength limit and Warren-Cowley short range order parameter

Concentration-concentration fluctuations in the long-wavelength limit ($Scc(0)$) is a microscopic function that is useful in the study of the nature of atomic order. It provides information that determines the phase-separating or compound-forming nature of alloys. The $Scc(0)$ is related to $G_M$ and thermodynamic activities, as defined in Eq. (30). The measured $Scc(0)$ values are compared with their ideal values ($Scc^{id}(0)$) (Eq. (31)) at specified compositions to make useful deductions. When $Scc(0) < Scc^{id}(0)$ for a certain composition, it implies a tendency for heterocoordination, while $Scc(0) > Scc^{id}(0)$ implies a tendency for homocoordination.

$$Scc(0) = RT / \left(\frac{\partial^2 G_M}{\partial c_{Bi}^2}\right)_{T,P,N} = c_{Sn} a_{Bi} / \left(\frac{\partial a_{Bi}}{\partial c_{Bi}}\right)_{T,P,N} = c_{Bi} a_{Sn} / \left(\frac{\partial a_{Sn}}{\partial c_{Sn}}\right)_{T,P,N} \quad (30)$$

$$Scc^{id}(0) = c_{Bi} c_{Sn} \quad (31)$$

The second-order differential in Eq. (30) can be solved numerically to obtain $Scc(0)$.

The Warren-Cowley short range order parameter, $α_1$, is used to determine the extent of order in the liquid alloy [34,35]. The $α_1$ can be deduced from the knowledge of the concentration-concentration and number-number structure factors from diffraction experiments. However, experimental structure factors are not easily measured during diffraction. Furthermore, $α_1$ can be determined from the knowledge of $Scc(0)$ and $Scc^{id}(0)$ as defined in Eq. (35) [34–37]. When $α_1 > 0$, it implies like-atom pairing as nearest neighbours, whereas $α_1 < 0$ corresponds to unlike-atom pairing as nearest neighbours, while $α_1 = 0$ denotes random distribution of atoms. The limiting values for $α_1$ when $c_{Bi} \leq 1/2$ and $c_{Bi} \geq 1/2$ are given in Eqs. (32) and (33), respectively. Equations (32) and (33) reduces to Eq. (34) when $c_{Bi} = c_{Sn} = 1/2$ [36].

$$-\frac{c_{Bi}}{c_{Sn}} \leq \alpha_1 \leq 1 \quad (32)$$

$$-\frac{c_{Sn}}{c_{Bi}} \leq \alpha_1 \leq 1 \quad (33)$$

$$-1 \leq \alpha_1 \leq +1 \quad (34)$$

When the value of $α_1$ is maximum (+1), it implies complete phase separation of components in the mixture, while its minimum value (-1) implies complete ordering of unlike-atoms as nearest neighbours. The relationship between $α_1$, $Scc(0)$ and $Scc^{id}(0)$ are provided in Eq. (35) [34–37].

$$\alpha_1 = \frac{\left(Scc(0)/Scc^{id}(0)\right)-1}{Scc(0)/Scc^{id}(0)\left(\hat{Z}-1\right)+1} \quad (35)$$

The Average coordination number, $\hat{Z}$, used in Eq. (35) was computed using Eq. (3).

## 2.3 Surface concentration and surface tension

A statistical mechanical approach to the modelling of surface properties, using the concept of a layered structure near the interface is known to be useful in binary alloys [27,38,39]. The grand partition functions setup for the surface layer and the bulk provides a relation between surface and bulk compositions, given in Eq. (36) [40]. In Eq. (36), $σ$ is the surface tension of the mixture, $c_i^s$, $σ_i$, $γ_i$ and $γ_i^s$, are the surface concentration, surface tension, bulk activity coefficient and surface activity coefficient of component $i$ at temperature $T$, respectively, $A_0$ is the mean surface area of the mixture (defined in Eq. (37)), $N_0$ is Avogadro's number and $k_B$ is Boltzmann's constant. The surface activity coefficient, $γ_i^s$, is defined in Eq. (38), where $γ_i(c_i^s)$ implies the use of $c_i^s$ in place of $c_i$ in the computation of activity coefficient. Similar to bulk properties (where $a_i = c_i \times γ_i$), surface activity ($a_i^s$) is a product of $c_i^s$ and $γ_i^s$.

Equation (36) is solved numerically to obtain surface concentration and surface tension for Bi-Sn at 600 and 900K.

$$\sigma = \sigma_{Bi} + \frac{k_B T}{A_0}\ln\frac{c_{Bi}^s}{c_{Bi}} + \frac{k_B T}{A_0}\ln\frac{\gamma_{Bi}^s}{\gamma_{Bi}} = \sigma_{Sn} + \frac{k_B T}{A_0}\ln\frac{c_{Sn}^s}{c_{Sn}} + \frac{k_B T}{A_0}\ln\frac{\gamma_{Sn}^s}{\gamma_{Sn}} \quad (36)$$

$$A_0 = 1.102 N_0^{-2/3}\left[c_{Bi} V_{mBi}^{2/3} + c_{Sn} V_{mSn}^{2/3}\right] \quad (37)$$

$$\ln \gamma_i^s = p \ln \gamma_i\left(c_i^s\right) + q \ln \gamma_i \quad (38)$$

In Eq. (38), $p$ and $q$ are surface coordination functions such that $p + 2q = 1$ ($p$ & $q$ are $1/2$ & $1/4$, respectively, for closed packed structures). Where $p$ and $q$ can be defined as fractions of the total number of nearest neighbours made by an atom within the layer in which it lies and that in the adjoining layer, respectively.

**Table 3** gives the surface tension for pure Bi & Sn at 600 & 900K, and these values are applied in the solution of Eq. (36).

**Tab. 3:** Surface tension for pure components.

| Metal | $\sigma_i$ [a] (N m$^{-1}$) | |
| --- | --- | --- |
| l | 600K | 900K |
| Bi | 0.3741 | 0.3531 |
| Sn | 0.5515 | 0.5245 |

[a] Obtained from [26]

# 3 Results and discussion

Based on the theoretical formalism described in Section 2, it is important to note that the fitted interaction parameters will remain unchanged in the calculations of enthalpy and entropy of formation for Bi-Sn at 600K; concentration fluctuations, Warren-Cowley short range order parameter, surface concentrations and surface tension at 600 and 900K. The fitted parameters will form the basis for understanding the energetics of the alloys.

## 3.1 Enthalpy and entropy of mixing

The partial differentials of interaction parameters for Bi-Sn at 600K were fitted while retaining the interaction parameters presented in **Tab. 2**. Satisfactory partial differentials, presented in **Tab. 4**, give good predictions of enthalpy and entropy mixing ($H_M$ and $S_M$) [10]. Parameter $g$ in **Tab. 2** agrees with $\partial g/\partial T$ (**Tab. 4**), in the sense that the negative $\partial g/\partial T$ suggests decreased energy of formation of chemical complexes with temperature increase. The positive $\partial v_i s/\partial T$ values can be interpreted as increasing repulsion between pseudo-ternary species as system temperature increases. Conversely, negative $\partial v_i s/\partial T$ values can be interpreted as decreasing repulsion between pseudo-ternary species as system temperature increases. The agreements between the theoretical fitting and experimentally observed values of $H_M$ and $S_M$ were quantified by estimating $Er_i$ (Eq. (23)). Excess entropy of mixing ($S_M^{xs}$) was estimated by applying Eqs. (27) & (29).

**Tab. 4:** Partial differentials of interaction parameters obtained from experimental data [10].

| System | $\dfrac{\partial g}{\partial T}$ | $\dfrac{\partial v_{12}}{\partial T}$ | $\dfrac{\partial v_{13}}{\partial T}$ | $\dfrac{\partial v_{23}}{\partial T}$ | $Er_i$ (%) | |
| --- | --- | --- | --- | --- | --- | --- |
| | | | | | $H_M$ | $S_M$ |
| Bi-Sn at 600K | -15.0 R | 2.6 R | -0.1 R | 5.2 R | 2.90 | 0.23 |

Furthermore, $H_M/RT$ and $S_M^{xs}/R$ for all concentration range were estimated and the results are graphically represented with plots in **Fig. 3**. The $Er_i$ values, presented in **Tab. 4**, and the $H_M/RT$ and $S_M^{xs}/R$ plots in **Fig. 3** show good agreements between the fitted theoretical values and experiments [10].

The $H_M/RT$ plot in **Fig. 3** shows that Bi-Sn at 600K is symmetric about the equiatomic composition and exhibits positive deviation from Raoultian behaviour across all concentration range. Also, the $S_M^{xs}$ plot in **Fig. 3** reveal negative values in all composition range and asymmetry around the equiatomic composition.

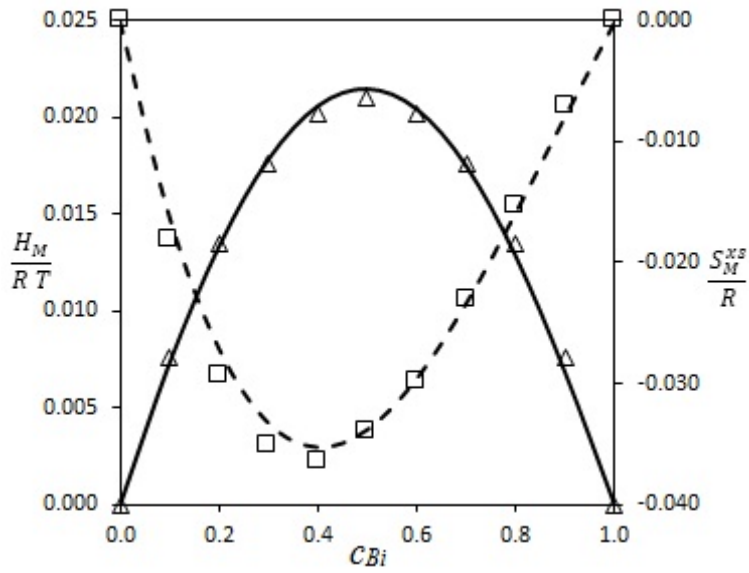

**Fig. 3:** Concentration dependence of $H_M/RT$ and $S_M^{xs}/R$ at 600K. Triangles represent $H_M/RT$ obtained from experiments [10], squares represent $S_M^{xs}/R$ obtained from experiments [10], solid lines represent QLT-computed $H_M/RT$ while dashed lines represent QLT-computed $S_M^{xs}/R$.

## 3.2 Concentration-concentration fluctuations in the long-wavelength limit and Warren-Cowley short range order parameter

The $Scc(0)$ for Bi-Sn at 600 and 900K, were computed and the results were plotted and presented in **Fig. 4**. The interpretation of the $Scc(0)$ results depends on the corresponding ideal values at every composition. **Figure 4** shows homocoordination tendency in the entire composition range of Bi-Sn at 600 and 900K.

The $\alpha_1$ for the Bi-Sn systems were computed and the results were plotted and presented in **Fig. 5**. **Figure 5** shows that $\alpha_1 > 1$ for all compositions and at both temperatures, which corroborates the $Scc(0)$ results. The peak value of $\alpha_1$ is about $+0.015$, this implies that a slight homocoordination property exists in the Bi-Sn systems, in other words, there is a slight preference for like-atoms as nearest neighbours. It can also be observed that, at about $c_{Bi} \geq 0.35$, segregation tendency increases with a rise in temperature, while the reverse is the case at about $c_{Bi} < 0.35$.

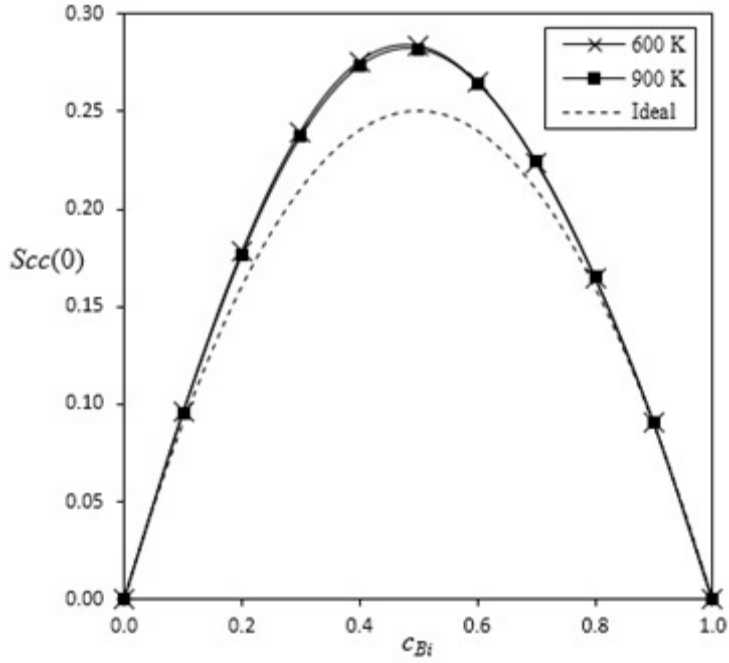

**Fig. 4:** Scc(0) for Bi-Sn at 600 & 900K.

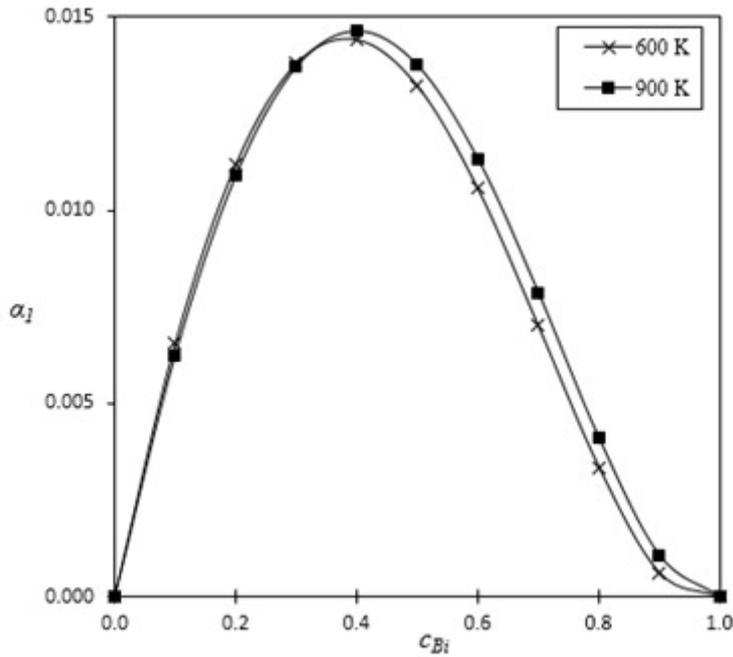

**Fig. 5:** α1 for Bi-Sn at 600 & 900K.

### 3.3 Surface concentration and surface tension

The surface concentrations of Bi & Sn, as well as the surface tension of the Bi-Sn system, were obtained simultaneously, as functions of bulk concentrations by solving Eq. (36), and the results are presented as plots in **Fig. 6** and **Fig. 7**. The plots in **Fig. 6** show an expected trend, particularly, the surface concentration increases with an increase in bulk concentration for Bi and Sn at 600 and 900K. Also, **Fig. 6** shows that the surface of the Bi-Sn system is enriched with Bi-atoms within 600 and 900K, for example, when $c_{Bi} \approx 0.1$, $c_{Bi}^s \approx 0.4$ at 900K and $c_{Bi}^s \approx 0.6$ at 600K. The surface concentration of Bi decreases as temperature of Bi-Sn rises. Since $c_{Bi}^S + c_{Sn}^S = 1$, it can also be deduced that the surface concentration of Sn increases as temperature of Bi-Sn rises.

The computed surface tension of Bi-Sn at 600 and 900K, presented in **Fig. 7**, reduces as temperature rises. This surface tension plots have a concave-like shape which agrees with many studies on surface properties that the component with lower surface tension, that is, Bismuth, has the greatest surface enrichment.

Due to the lack of known experimental data, the surface concentrations and surface tension calculations could not be subjected to comparison.

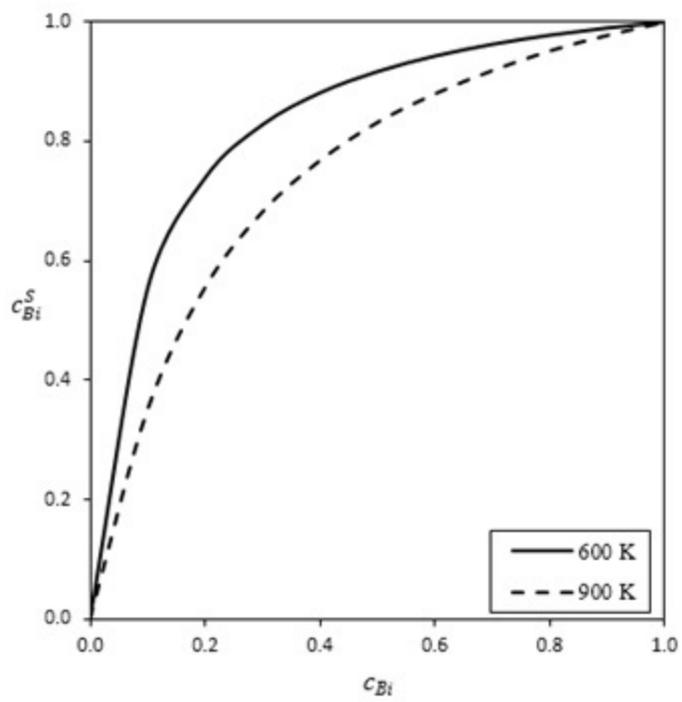

**Fig. 6:** $c_{Bi}^S$ vs. $c_{Bi}$ for Bi-Sn at 600 & 900K.

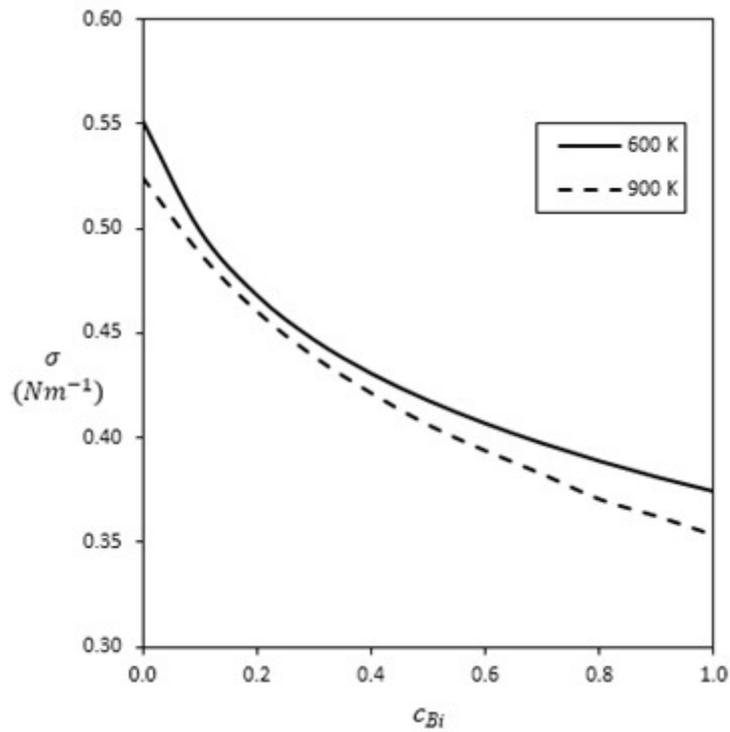

**Fig. 7:** Concentration dependence of surface tension for Bi-Sn at 600 & 900K.

# 4 Conclusions

The improved QLT [15], which applies nature, composition, and temperature-dependent coordination numbers, has been successfully applied to model the thermodynamic properties of Bi-Sn at 600 and 900K.

The fitted parameters were observed to reliably predict the Gibbs free energies of mixing and thermodynamic activities of alloy components. The same set of interaction parameters were used while fitting their temperature derivatives, to model the integral enthalpies and entropies of mixing of Bi-Sn at 600K. The integral enthalpies of mixing exhibited positive deviation from Raoultian behaviour and is symmetric about the equiatomic composition, while the integral excess entropies of mixing is negative in all composition range and asymmetric about the equiatomic composition. The $Scc(0)$ and $\alpha_1$ computations for Bi-Sn at 600 and 900K revealed slight segregation tendencies. Surface property studies reveal that Bi atoms segregate to the Bi-Sn surface, and the enrichment of Bi atoms at the surface reduces with temperature. The surface tension for Bi-Sn at 600 and 900K was also predicted. The information provided in this work will help in the prediction of bulk and surface properties of Bi-Sn-based ternary and multicomponent systems.

# References

[1] M. Abtew, G. Selvaduray, Lead-free solders in microelectronics, Mater. Sci. Eng. R Reports. 27 (2000) 95–141. https://doi.org/10.1016/S0927-796X(00)00010-3.

[2] N.A. Asryan, A. Mikula, Thermodynamic properties of Bi-Sn Melts, Inorg. Mater. 40 (2004) 386–390.

[3] N. Moelans, K.C.H. Kumar, P. Wollants, Thermodynamic optimization of the lead-free solder system Bi–In–Sn–Zn, Calphad Comput. Coupling Phase Diagrams Thermochem. 360 (2003) 98–106.

[4] O.E. Awe, O.M. Oshakuade, Theoretical prediction of thermodynamic activities of liquid Au-Sn-X (X=Bi, Sb, Zn) solder systems, Phys. B Condens. Matter. 507 (2017) 84–94. https://doi.org/10.1016/j.physb.2016.11.035.

[5] O. Akinlade, F. Sommer, Concentration fluctuations and thermodynamic properties of ternary liquid alloys, J. Alloys Compd. 316 (2001) 226–235. https://doi.org/10.1016/S0925-8388(00)01450-X.

[6] K.C. Chou, S.K. Wei, A new generation solution model for predicting thermodynamic properties of a multicomponent system from binaries, Metall. Mater. Trans. B. 28 (1997) 439–445. https://doi.org/10.1007/s11663-997-0110-7.

[7] D.P. Tao, A new model of thermodynamics of liquid mixtures and its application to liquid alloys, Thermochim. Acta. 363 (2000) 105–113. https://doi.org/10.1016/S0040-6031(00)00603-1.

[8] I. Katayama, D. Živković, D. Manasijević, T. Tanaka, Ž. Živković, H. Yamashita, Thermodynamic Properties of Liquid Sn-Bi-Sb Alloys, Netsu Sokutei. 32 (2005) 40–44. https://doi.org/10.11311/jscta1974.32.40.

[9] O.E. Awe, O.M. Oshakuade, Theoretical prediction of thermodynamic activities of all components in the Bi-Sb-Sn ternary lead-free solder system and Pb-Bi-Sb-Sn quaternary system, Thermochim. Acta. 589 (2014) 47–55. https://doi.org/10.1016/j.tca.2014.05.009.

[10] R. Hultgren, P.D. Desai, D.T. Hawkins, M. Geiser, K.K. Kelley, eds., Selected values of the thermodynamic properties of binary alloys, ASM, Metals Park, OH., 1973.

[11] D. Manasijević, D. Minić, D. Živković, I. Katayama, J. Vřešťál, D. Petković, Experimental investigation and thermodynamic calculation of the Bi-Ga-Sn phase equilibria, J. Phys. Chem. Solids. 70 (2009) 1267–1273. https://doi.org/10.1016/j.jpcs.2009.07.010.

[12] I. Katayama, T. Tanaka, S. Akai, K. Yamazaki, T. Iida, Activity Measurement of Liquid Sn-Ag-Bi Alloys by Fused Salt EMF Method, Mater. Sci. Forum. 502 (2005) 129–138. https://doi.org/10.4028/www.scientific.net/msf.502.129.

[13] A.B. Bhatia, R.N. Singh, A quasi-lattice theory for compound forming molten alloys, Phys. Chem. Liq. 13 (1984) 177–190. https://doi.org/10.1080/00319108408080778.

[14] A.B. Bhatia, W.H. Hargrove, Concentration fluctuations and thermodynamic properties of some compound forming binary molten systems, Phys. Rev. B. 10 (1974) 3186. https://doi.org/10.1103/PhysRevB.10.3186.

[15] Oshakuade OM, Awe OE. Modification of the quasi-lattice theory for liquid alloys on the basis of varying the coordination number and its application to Al-Sn, Al-Zn and Sn-Zn. Submitted for publication. n.d. arXiv:2102.08516 [cond-mat.mtrl-sci].

[16] D.P. Tao, Prediction of the coordination numbers of liquid metals, Metall. Mater. Trans. A. 36 (2005) 3495–3497. https://doi.org/10.1007/s11661-005-0023-5.

[17] C. Kittel, Introduction to Solid State Physics, 7th ed., John Wiley & Sons Inc., New York, 1996.

[18] F.H. Trimble, N.S. Gingrich, The effect of temperature on the atomic distribution in liquid sodium, Phys. Rev. 53 (1938) 278–281. https://doi.org/10.1103/PhysRev.53.278.

[19] N.S. Gingrich, L. Heaton, Structure of Alkali Metals in the Liquid State, J. Chem. Phys. 34 (1961) 873–878. https://doi.org/10.1063/1.1731688.

[20] P.C. Sharrah, J.I. Petz, R.F. Kruh, Determination of atomic distributions in liquid lead-bismuth alloys by neutron and x-ray diffraction, J. Chem. Phys. 32 (1960) 241–246. https://doi.org/10.1063/1.1700908.

[21] J.R. Wilson, The structure of liquid metals and alloys, Metall. Rev. 10 (1965) 381–590. https://doi.org/10.1179/mtlr.1965.10.1.381.

[22] A.L. Hines, H.A. Walls, K.R. Jethani, Determination of the coordination number of liquid metals near the melting point, Metall. Trans. A. 16 (1985) 267–274. https://doi.org/10.1007/BF02815308.

[23] J.R. Cahoon, The first coordination number for liquid metals, Can. J. Phys. 82 (2004) 291–301. https://doi.org/10.1139/p04-003.

[24] S.R. Elliott, Physics of amorphous materials, Longman House, Essex, 1983.

[25] G. Saffarini, Glass transition temperature and molar volume versus average coordination number in $Ge_{100-x}S_x$ bulk glasses, Appl. Phys. A Solids Surfaces. 59 (1994) 385–388. https://doi.org/10.1007/BF00331716.

[26] T. Iida, R.I.L. Guthrie, The physical properties of liquid metals, Clarendon Press, Oxford, 1988.

[27] E.A. Guggenheim, Mixtures, Oxford University Press, London, 1952.

[28] J.M. Prausnitz, R.N. Lichtenthaler, E.G. de Azevedo, Molecular thermodynamics of fluid-phase equilibria, 2nd ed., Prentice-Hall Inc., New Jersey, 1986.

[29] R. Novakovic, T. Tanaka, Bulk and surface properties of Al-Co and Co-Ni liquid alloys, Phys. B Condens. Matter. 371 (2006) 223–231. https://doi.org/10.1016/j.physb.2005.10.111.

[30] U. Khair, H. Fahmi, S. Al Hakim, R. Rahim, Forecasting Error Calculation with Mean Absolute Deviation and Mean Absolute Percentage Error, in: J. Phys. Conf. Ser., Institute of Physics Publishing, 2017. https://doi.org/10.1088/1742-6596/930/1/012002.

[31] O. Akinlade, Thermodynamics of molten K-Te alloys, J. Phys. Condens. Matter. 6 (1994) 4615. https://doi.org/10.1088/0953-8984/6/25/001.

[32] O. Akinlade, Ordering phenomena in Na-Ga and Na-Sn molten alloys, Phys. Chem. Liq. 29 (1995) 9–21. https://doi.org/10.1080/00319109508030260.

[33] N.A. Gokcen, Statistical thermodynamics of alloys, Plenum Press, New York, 1986.

[34] B.E. Warren, X-Ray Diffraction, Addison-Wesley, Reading, 1969.

[35] J.M. Cowley, An approximate theory of order in alloys, Phys. Rev. 77 (1950) 669–675. https://doi.org/10.1103/PhysRev.77.669.

[36] R.N. Singh, Short-range order and concentration fluctuations in binary molten alloys, Can. J. Phys. 65 (1987) 309–325. https://doi.org/10.1139/p87-038.

[37] R.N. Singh, D.K. Pandey, S. Sinha, N.R. Mitra, P.L. Srivastava, Thermodynamic properties of molten LiMg alloy, Phys. B+C. 145 (1987) 358–364. https://doi.org/10.1016/0378-4363(87)90105-7.

[38] O. Akinlade, R.N. Singh, Bulk and surface properties of liquid In–Cu alloys, J. Alloys Compd. 333 (2002) 84–90. https://doi.org/10.1016/S0925-8388(01)01733-9.

[39] O. Akinlade, O.E. Awe, Bulk and surface properties of liquid Ga-Tl and Zn-Cd alloys, Int. J. Mater. Res. 97 (2006) 377–381. https://doi.org/10.3139/146.101227.

[40] L.C. Prasad, R.N. Singh, G.P. Singh, The Role of Size Effects on Surface Properties, Phys. Chem. Liq. 273 (1994) 179–185. https://doi.org/10.1080/00319109408029523.